\documentclass[11pt]{article}
\pdfoutput=1
\usepackage{graphics}
\usepackage[left=1in,top=1in,right=1in,bottom=1in]{geometry}
\usepackage{mhchem}
\usepackage{upgreek}
\usepackage{hyperref}   
\usepackage{authblk}   
\usepackage{caption}

\begin{document}
\title{Active colloidal particles in emulsion droplets:\\ A model system for the cytoplasm}
\author[1,2]{Viva R.\ Horowitz} 
\author[1]{Zachary C.\ Chambers}
\author[3,4,5,6]{\.{I}rep G\"{o}zen}
\author[1]{Thomas G.\ Dimiduk}
\author[3,1]{Vinothan N.\ Manoharan\footnote{\href{mailto:vnm@seas.harvard.edu}{vnm@seas.harvard.edu}}}
\affil[1]{Department of Physics, Harvard University, Cambridge MA
  02138, USA}
\affil[2]{ Department of Physics, Hamilton College, Clinton NY
  13323 USA} 
\affil[3]{Harvard John A.\ Paulson School of Engineering and
  Applied Sciences, Harvard University, Cambridge MA 02138 USA}
\affil[4]{Faculty of Medicine, University of Oslo, 0318 Oslo Norway}
\affil[5]{Faculty of Mathematics and Natural Sciences, University of Oslo, 0315 Oslo,
  Norway}
\affil[6]{Department of Chemistry and Chemical Engineering, Chalmers
  University of Technology, SE-412 96 G\"{o}teborg, Sweden}
\maketitle
{\abstract{In living cells, molecular motors create activity that
  enhances the diffusion of particles throughout the cytoplasm, and not
  just ones attached to the motors. We demonstrate initial steps toward
  creating artificial cells that mimic this phenomenon. Our system
  consists of active, Pt-coated Janus particles and passive tracers
  confined to emulsion droplets. We track the motion of both the active
  particles and passive tracers in a hydrogen peroxide solution, which
  serves as the fuel to drive the motion. We first show that correcting
  for bulk translational and rotational motion of the droplets induced
  by bubble formation is necessary to accurately track the particles.
  After drift correction, we find that the active particles show
  enhanced diffusion in the interior of the droplets and are not
  captured by the droplet interface. At the particle and hydrogen
  peroxide concentrations we use, we observe little coupling between the
  active and passive particles. We discuss the possible reasons for lack
  of coupling and describe ways to improve the system to more
  effectively mimic cytoplasmic activity.}}
\section{Introduction}
\label{intro}
The interior of a biological cell is an active environment driven by the
motion of motor proteins. These molecular motors not only transport
their cargo, but also enhance the transport of unattached particles in
the cytoplasm~\cite{Guo14,Crocker07}. The stresses created by the motors
cause such particles to diffuse at a rate higher than that expected from
thermal fluctuations alone. Indeed, in ATP-depleted cells, the
mean-square displacement (MSD) of unattached tracer particles is nearly
an order of magnitude smaller than that in untreated cells at lag times
of 10~s or more~\cite{Guo14}, indicating that even particles not
directly transported by motors experience an increase in motion. These
results suggest that motors play an important role in driving and
dispersing a wide range of cellular components.

To better understand how active components can drive unattached, passive
particles in confined systems, we study a minimal model for an
artificial cell, consisting of colloidal Janus particles and passive
particles inside a water-in-oil emulsion droplet. The Janus particles
are coated with platinum on one hemisphere, so that they catalyze the
breakdown of hydrogen peroxide~\cite{Howse07}:
\begin{equation}
\ce{2H2O2 ->[Pt] 2H2O + O2}.
\end{equation}
The resulting gradient of dissolved oxygen leads to
self-diffusiophoresis~\cite{Howse07,wang_selecting_2014}, providing one
mechanism for the Janus particle to move, although self-electrophoresis
is likely also important~\cite{brown_ionic_2014}. We observe and track
both the active and passive particles to understand how the motions of
the two are coupled. As in the cellular system, the activity can be
controlled by depleting the system of fuel---in this case, the hydrogen
peroxide.

After correcting for drift induced by the production of oxygen bubbles
and, to a lesser extent, by sample evaporation, we find that the active
particles do show enhanced diffusion in the emulsion droplet, but the
passive particles do not diffuse any faster than in the absence of the
fuel. The absence of coupling gives insights into what physical
considerations may be important for the cell to convert active motion
into enhanced diffusion.

\section{Methods and materials}
\label{sec:methods}

To make droplets containing Janus particles, tracer particles, and
hydrogen peroxide fuel, we first make two batches of dyed polystyrene
particles, each with a different dye. One batch is then coated with
platinum on one hemisphere to make the Janus particles, and the other
becomes the tracer particles. We then mix the particles and encapsulate
them in water droplets containing hydrogen peroxide. We use
1-$\upmu$m-diameter particles in all of our experiments because they are
large enough to track easily but small enough that they do not sediment
rapidly.

Next, we use fluorescence microscopy to observe the motion of the
particles inside the droplet, taking advantage of the different dyes to
separate the signals from the Janus and tracer particles. Finally, we
track the particles and correct for bulk droplet motion in
post-processing.  The following subsections give details of each of
these steps.

\subsection{Dying particles}

To make both the passive tracer particles and Janus particles, we start
with spherical, sulfate polystyrene particles that are 1~$\upmu$m in
diameter (Invitrogen S37498) and fluorescently dye them so that we can
distinguish the two types of particles when we eventually put them into
droplets. We first wash the particles by centrifuging the solution,
removing the supernatant, and replacing it with an equal volume of
deionized water (Millipore) five times. We use green fluorescent dye
(BODIPY 493/503, Life Technologies, now Thermo Scientific) for the
particles that will become the Janus particles and red fluorescent dye
(Pyrromethene 650, Exciton) for the tracer particles. In both cases, we
saturate a toluene solution with the dye, add 4~$\upmu$L of dyed toluene
to 40~$\upmu$L of the suspension of particles at 1\%~w/v, and let the
particles swell on a room-temperature tube rotator overnight. We then
open the container to atmosphere and allow the toluene to evaporate in a
90$^\circ$C oven for 12~min. Finally, we wash the particles five times by
centrifugation and redispersion in deionized water.

\begin{figure}
\begin{center}
  \includegraphics{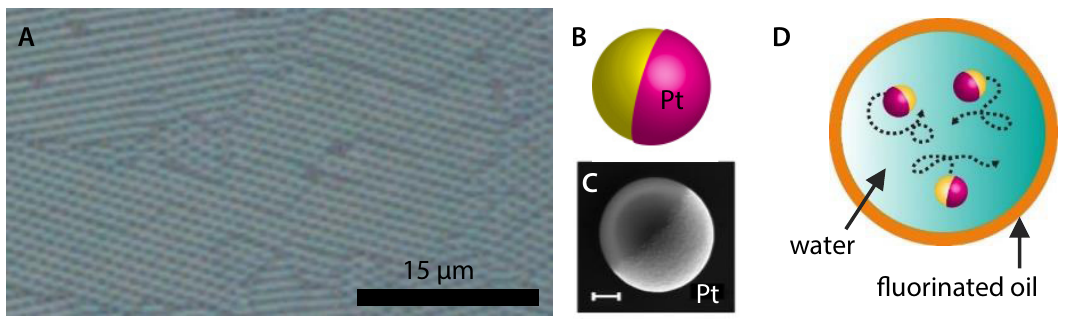}
  \caption{We spread polystyrene sulfate latex microspheres in a
    monolayer on a glass plate (A: brightfield optical micrograph), then
    evaporate platinum in a thin layer onto one hemisphere to create
    Janus particles (B: cartoon of a Janus particle, C: scanning
    electron micrograph obtained after drop-casting the suspension onto
    a silicon substrate. Scalebar is 200~nm). D: Simplified cartoon
    showing the active motion of Janus particles in a water droplet,
    which is typically 20--80 $\upmu$m in diameter. Our droplets also
    contain passive tracer particles, which are not shown here.}
\label{fig:diagram}  
\end{center}     
\end{figure}

\subsection{Making Janus particles}

We make Janus particles by coating half of each particle with a layer of
platinum to act as a catalyst for hydrogen peroxide. To do this, we
first dilute the suspension of green-dyed particles in ethanol to obtain
particles in 50\%~v/v ethanol/water. We then use a paint drawbar (BYK
bar film applicator, item 5552, with a clearance gap of 50.8~$\upmu$m)
to spread the suspension onto a 4$\times$5-in glass plate, and we
let the suspension dry at room temperature, after which we confirm by
microscopic inspection that the particles are in a monolayer on the
plate (Figure~\ref{fig:diagram}A). We then deposit 2~nm titanium (as an
adhesion layer) and 10~nm platinum onto the particles on the dried glass
plate using an electron-beam evaporator (Denton Explorer). The spherical
particles act as their own shadow mask, such that the metal is deposited
only onto the top hemisphere of each particle, creating the
characteristic two-faced particle (Figure~\ref{fig:diagram}B--C).

\subsection{Making an aqueous suspension of Janus particles}

To make the suspension of Janus and tracer particles, we first place the
glass plate in a plastic zip-sealed bag filled with deionized water and
sonicate it to release the particles, yielding a colloidal suspension of
Janus particles in about 1~L of water. We increase the concentration of
particles in suspension by centrifuging and removing the supernatant,
decreasing the total volume to 2 or 3~mL. To estimate the concentration
of the colloidal suspension, we compare microscope images of the Janus
particles in water to images of polystyrene particles at 1\%~w/v and
adjust the concentration until the number of particles per volume is
approximately matched. We then mix the green Janus particles and red
tracer particles in deionized water to obtain a solution with 0.5\%~w/v
tracer particles and a comparable number density of Janus particles. The
final, total number density of particles is approximately the same as
that of a 1\%~w/v solution of polystyrene particles. We assume that this
number density is the same as that inside the droplets, which are
prepared as described below.

\subsection{Encapsulating Janus particles and tracer particles together in emulsion droplets}

We form aqueous droplets containing Janus particles, tracers, and
hydrogen peroxide fuel. Timing is a factor in this step because once the
peroxide comes into contact with the platinum on the Janus particles, it
starts to break down into water and oxygen. The chemical reaction
continues until the hydrogen peroxide is exhausted. To observe the
system while the reaction is in progress, we image it within a few
minutes of preparation.

We prepare the droplets by first adding hydrogen peroxide (Electron
Microscopy Sciences, 30\%~v/v) to the suspension of Janus and tracer
particles to make up a solution at 3\%~v/v hydrogen peroxide. This
concentration is small enough to prevent rapid formation of oxygen
bubbles, but high enough to produce active motion. All of the samples we
discuss have either 0\% (unpowered) or 3\% (powered) hydrogen peroxide.
We then combine 15 $\upmu$L of this aqueous solution with 60~$\upmu$L of
fluorinated oil (HFE-7500, 3M Novec) containing 2\%~w/w surfactant
(008-FluoroSurfactant, Ran Biotechnologies, Beverly, MA) and vortex this
mixture for 30 to 60~s. The resulting water-in-oil emulsion contains
water droplets with diameters between about 20 and 80~$\upmu$m
(Figure~\ref{fig:diagram}D). We use this particular combination of
ingredients because the fluorinated oil and surfactant prevent the
particles from adhering to the interface of the
droplets~\cite{holtze_biocompatible_2008}, thus allowing them to move
about the droplet interior.

To observe the droplets under the microscope, we pipette 6~$\upmu$L of
the emulsion onto an open-top platform prepared by epoxying a metal
washer to a glass coverslip. The exposed top surface allows oxygen
bubbles from the chemical reaction to escape the sample. We then
immediately place the sample on an inverted microscope (Nikon TE2000)
with a 60$\times$ water-immersion objective (CFI Plan Apo, Nikon) and
1.5$\times$ tube lens for fluorescence imaging.

\subsection{Recording the particle motion}

\begin{figure}
\begin{center}
  \includegraphics{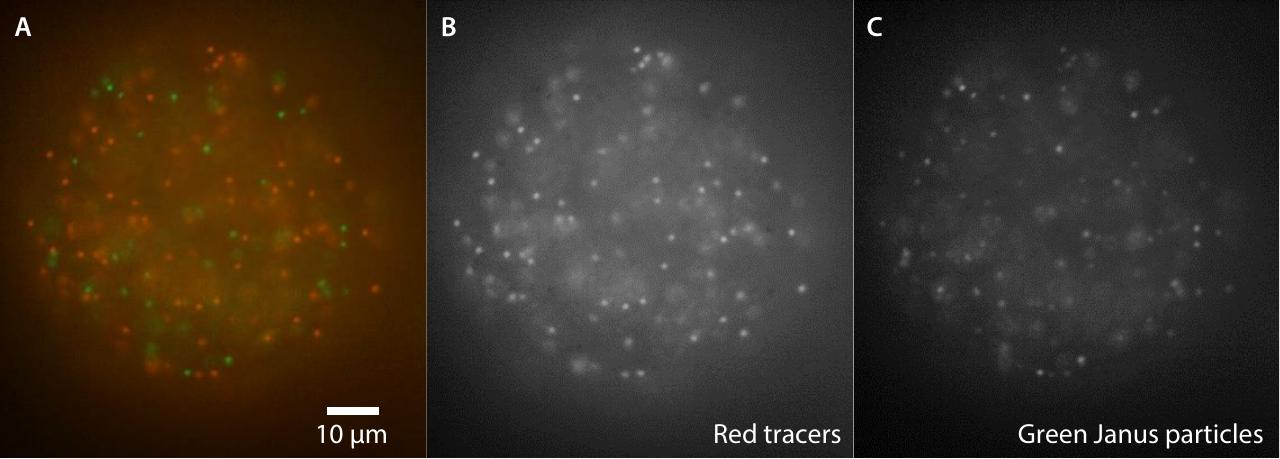}
  \caption{Red tracer particles and green Janus particles are
    encapsulated in an aqueous droplet (70 $\upmu$m diameter), which
    serves as our simplified model of an artificial cell (A). We analyze
    the red channel to track the tracer particles (B) and the green
    channel to track the Janus particles (C).
}
  \label{fig:images}
\end{center}
\end{figure}

Using the microscope chamber described above, we record fluorescence
images of Janus particles and passive tracer particles in the aqueous
droplets (Figure~\ref{fig:images}A). The aqueous droplets float in
fluorinated oil. We image through the oil to the lowest droplets,
furthest from the emulsion-air interface, where evaporation of the oil
and water can cause droplets to move. We fluorescently image the two
types of dyed particles simultaneously with an LED illuminator
(Spectra-X light engine, Lumencor, Beaverton, OR) and a multiband
filtercube (DA/FI/TR/Cy5-A-NTE, Semrock, Rochester, NY). We collect AVI
videos using a color camera (DCC3240C, Thorlabs, Newton, NJ) and
associated software. With this widefield technique, we observe only
those particles that are close to the focal plane.

\subsection{Tracking particle trajectories}

To analyze the motion of the two species of particles, we first use the
Fiji distribution of the ImageJ software
package~\cite{schindelin_fiji:_2012} to separate the images into red and
green files (Figure~\ref{fig:images}) based on the color channel of the
camera. Although some fluorescence bleeds into the opposite channel, it
is faint enough to remove with an appropriate choice of threshold. After
applying the threshold, we can clearly distinguish Janus particles from
tracer particles. We use the Python library Trackpy
(\url{https://soft-matter.github.io/trackpy}) \cite{trackpy} to locate
the centroids of the particles in each pre-processed frame and to
link the coordinates of these particles together into trajectories.

\subsection{Subtracting rotational and translational drift}

We calculate the overall drift of the droplets to separate the random
motion within the droplets from the coherent motion of the droplets
themselves. We use a singular value decomposition (SVD)
method~\cite{BeslMcKay} to model the three-dimensional rigid-body
rotation and translation of the droplet, based on the two-dimensional
particle trajectories. The rotation is calculated about a different axis
for each time step, and the rotational and translational drift is
subtracted from the trajectories to obtain the motion in the droplet
frame.

\section{Results}
\label{sec:results}

\subsection{Observations of system}
\label{sec:observations}

As shown by the microscope images in Figure~\ref{fig:images}, we are
able to create stable emulsion droplets with the two types of particles
and the hydrogen peroxide. The particles do not aggregate in the
interior of the droplets, and we do not observe them adhering to the
interface.

\subsection{Droplet motion affects particle trajectories}

\begin{figure}
  \begin{center}
    \includegraphics{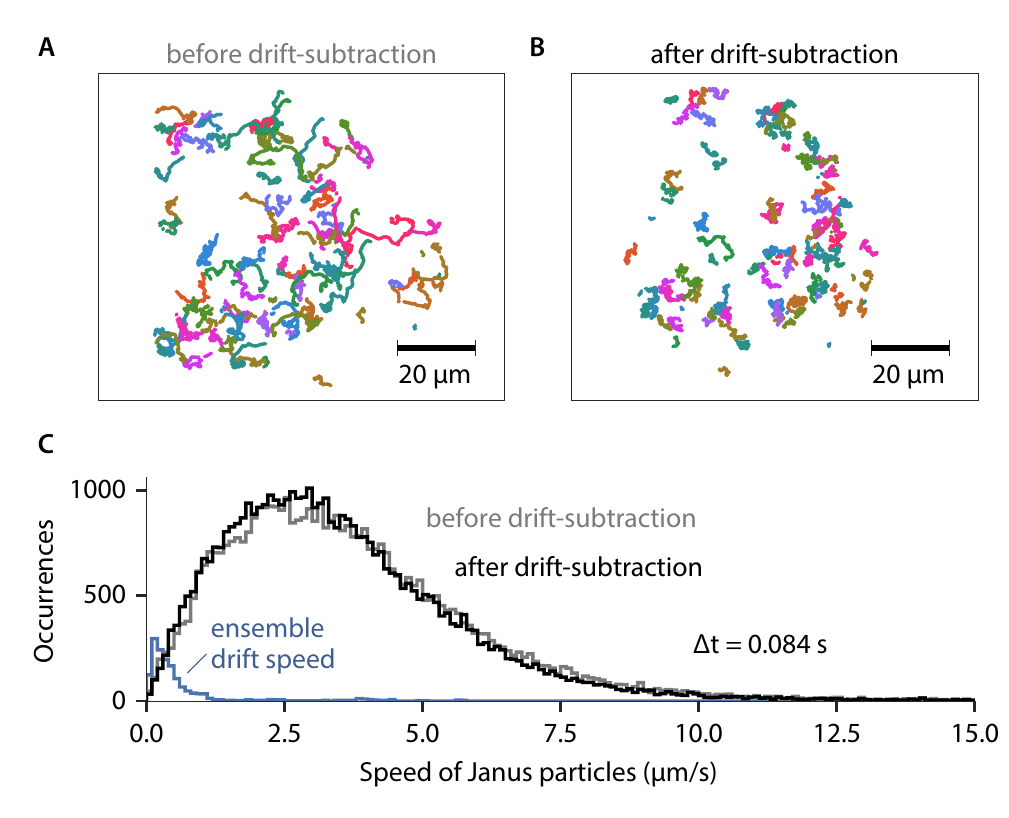}
    \caption{We track powered Janus particles in an emulsion droplet.
      Prior to any drift subtraction (A), the trajectories show evidence
      of correlated motion arising from translation and rotation of the
      droplet. After drift subtraction (B), there is less correlated
      motion, and we observe the particles exploring a smaller area.
      These trajectories are selected from the portion of the video
      sequence that showed the largest translational drift, so as to
      illustrate the effect clearly; in most cases the drift is not so
      easily discerned by eye. C: Speed distributions of Janus particles
      before (gray) and after (black) drift subtraction, compared to the
      distribution of translational drift speeds (blue). The plot shows
      that the Janus particles move faster than the ensemble-averaged
      drift speed. Thus, while drift subtraction removes the correlated
      motion of particles arising from movement of the droplet, we do
      not expect it to remove the powered motion of the Janus particles
      themselves. All speeds are measured in the focal plane of the
      droplet shown in Figure~\ref{fig:images}. The areas of the
      histograms differ because the number of data points for the
      ensemble is the number of frames in the trajectory, while the
      number of data points in the particle histograms is the number of
      frames times the number of particles in each frame. }
    \label{fig:trajectories}
  \end{center}
\end{figure}

We observe large-scale translational and rotational motion of our
droplets that most likely arises from growing oxygen bubbles. Although
translational drift is a common problem in particle tracking
experiments, the bulk rotational motion is especially problematic in the
droplet system, since each droplet can rotate independently. The effects
of the bulk motion can be seen in the trajectories of the Janus
particles in Figure~\ref{fig:trajectories}A, which show an elongation
that corresponds to a large-scale motion of the droplet, likely arising
from forces external to it.

To correct for this effect, we calculate the rotational and
translational drift as described in the Methods section, and we subtract
it from the particle trajectories. The resulting trajectories, shown in
Figure~\ref{fig:trajectories}B, explore a smaller area with less
elongation.

To determine whether the drift subtraction artificially reduces the
measured, powered motion of the Janus particles, we calculate the speed
distribution of the Janus particles before and after drift subtraction
(Figure~\ref{fig:trajectories}C). The speed is calculated from the
displacement $\Delta r$ in the focal plane measured over the interval
between frames and divided by the frame interval. The ensemble drift
speed is the ensemble-average displacement $\sqrt{\left< \Delta x
  \right>^2 +\left< \Delta y \right>^2 }$ for each frame, again divided
by the interval between frames. We see that the ensemble-average speed
is peaked at 0.2~$\upmu$m/s with a mean of 0.65~$\upmu$m/s and median of
0.34~$\upmu$m/s. The speed of the Janus particles is peaked at
2.5~$\upmu$m/s with a mean of 3.7~$\upmu$m/s and median of
3.3~$\upmu$m/s before subtraction and a mean of 3.6~$\upmu$m/s and
median of 3.2~$\upmu$m/s after subtraction. The drift correction
therefore does not significantly affect the measured motion of the
particles at short times, because the drift velocity is small compared
to the velocity of the particles.

\begin{figure}
\begin{center}
  \includegraphics{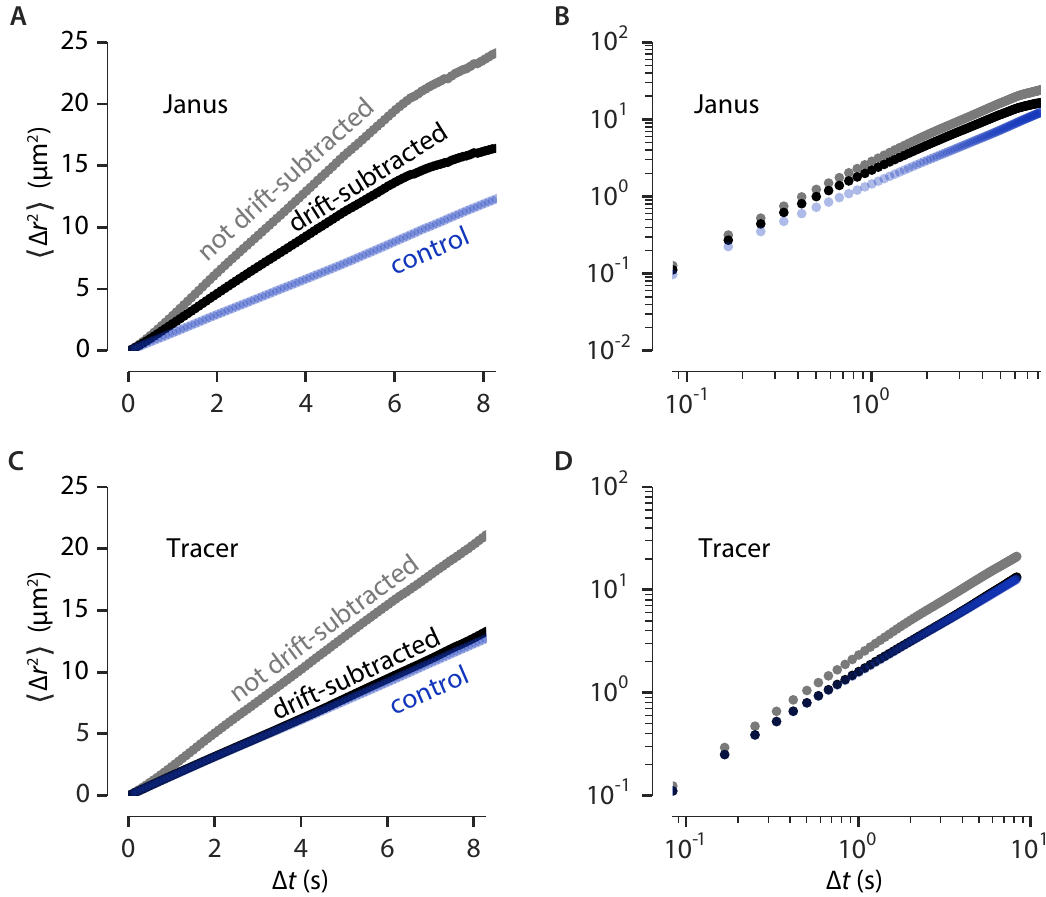}
  \caption{The ensemble mean square displacements (MSD) $\langle\Delta
    r^2 \rangle$ of both active Janus (A,B) and tracer particles (C,D)
    in the same droplet as a function of lag time $\Delta t$. The
    two-dimensional ensemble MSD is shown both before (gray) and after
    (black) drift-subtraction. The log-log plots at right (B,D) show the
    same datasets as the left (A,C). A control sample lacking hydrogen
    peroxide (blue) is shown for comparison.}
  \label{fig:ensemble-MSD}
\end{center}
\end{figure}

We also characterize the motion by calculating the ensemble-average MSD,
and we compare it to that measured in a control experiment, where we use
droplets containing both types of particles but do not add any hydrogen
peroxide (Figure~\ref{fig:ensemble-MSD}). For both Janus particles and
tracers, the drift subtraction reduces the apparent MSD at all lag
times, as expected. The MSD becomes smaller because the drift is not
random. Instead, the motion of the droplet is correlated over many
frames. This effect, if left uncorrected, would lead to a spuriously
large MSD for both the Janus particles and tracers.

The drift-corrected MSDs show two features that we discuss in the next
two subsections. First, the Janus particles have a higher MSD at all lag
times than they do in the absence of hydrogen peroxide
(Figure~\ref{fig:ensemble-MSD}A), and the MSD on a log-log plot has a
larger slope (Figure~\ref{fig:ensemble-MSD}B). This enhancement persists
even after drift subtraction. Second, the drift correction brings the
MSD of the tracer particles in hydrogen peroxide much closer to that of
the particles in the absence of hydrogen peroxide
(Figure~\ref{fig:ensemble-MSD}C--D).

\subsection{Janus particles show enhanced diffusion in droplets}

\begin{figure}
\begin{center}
  \includegraphics{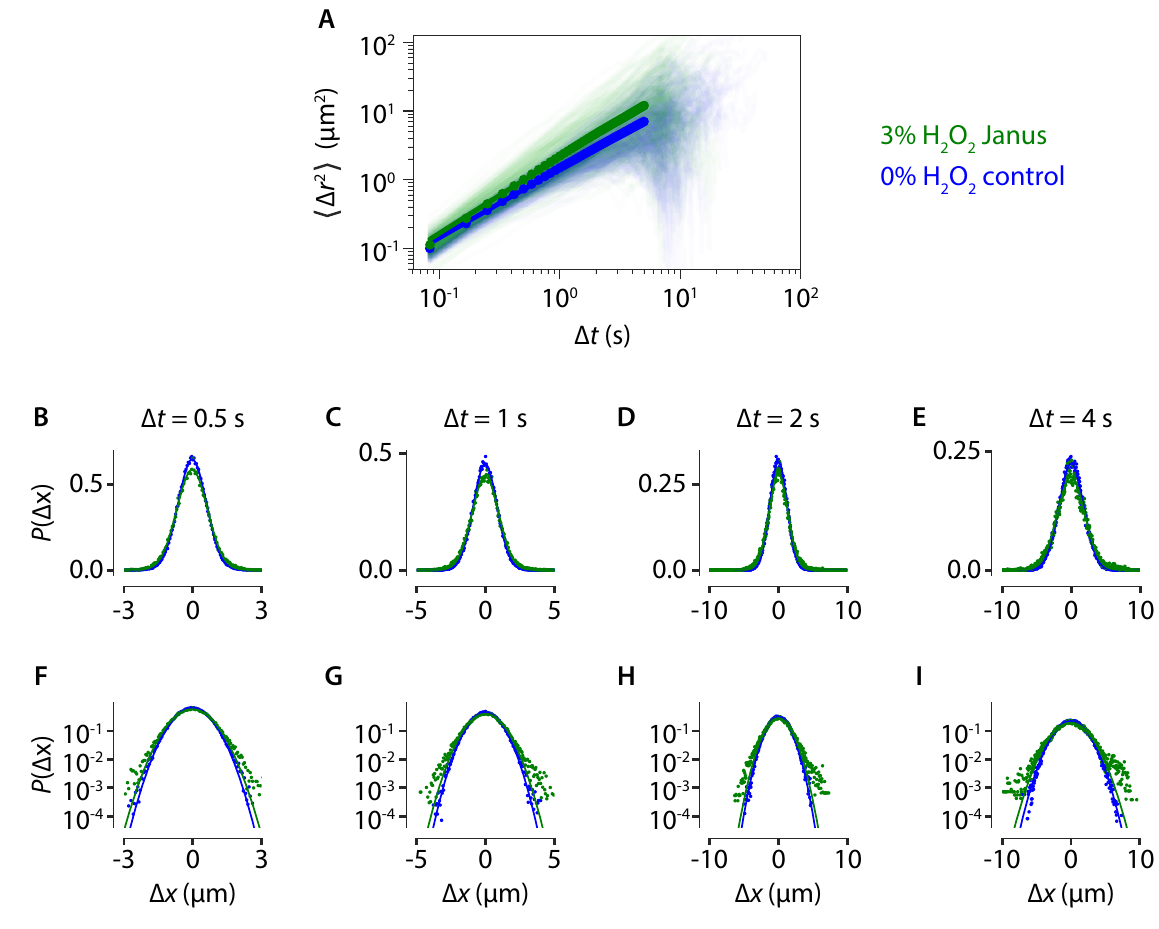}
  \caption{(Color online) Janus particles move superdiffusively in an
    emulsion droplet in the presence of hydrogen peroxide. A: Ensemble
    mean square displacement (MSD) as a function of lag time. The light
    traces show the MSD for individual particles, to indicate the spread
    of particle MSDs. The ensemble MSD log plot shows a greater slope
    for the powered Janus particles than the unpowered Janus particles,
    characteristic of superdiffusive motion. We fit the data up to
    8.4~s, because the noise is high at larger lag times, as shown by
    the spread of the traces (note that the noise appears uneven because
    the MSD is plotted on a log scale). B-D: Plots of the displacement
    distribution of powered (green) and unpowered (blue) Janus particles
    at lag times of 0.5~s (B), 1~s (C), 2~s (D), and 4~s (E). F--I: the
    same distributions shown on log-linear axes. The trajectories for
    these plots are drift-subtracted.}
  \label{fig:superdiffusion}
\end{center}
\end{figure}

The larger amplitude and slope of the MSD of the Janus particles (after
drift correction) relative to those of control particles suggest that
the particles are in fact being driven by the breakdown of the hydrogen
peroxide fuel. Further evidence for powered motion is shown in
Figure~\ref{fig:superdiffusion}, which shows the displacement
distributions for Janus particles as a function of lag time both with
and without hydrogen peroxide. Diffusive particles show Gaussian
distributions (Figure~\ref{fig:superdiffusion}B--E), as expected, with
widths corresponding to the ensemble mean square displacement
(Figure~\ref{fig:superdiffusion}A). The log plots
(Figure~\ref{fig:superdiffusion}F--I) reveal that the distribution of
the active Janus particles diverges from a Gaussian, with
larger-than-expected probability for a portion of the population of
Janus particles to travel more than 4 $\upmu$m in 1~s.

We first fit the MSD using a model of the form $\Delta r^2=A\Delta t^n$,
where $\Delta r^2$ is the two-dimensional MSD, $\Delta t$ is the lag
time, $A$ is a prefactor, and $n$ is an exponent.  We find
that for the unpowered particles $n=1.00\pm 0.03$, and for the powered
particles $n=1.10 \pm 0.05$, about 10\% higher. We estimate the
uncertainty in the fitting parameters by considering the variation among
individual particles. A typical standard error from the fit cannot be
used because the data at different lag times are correlated.

An exponent larger than 1.0, such as we observe here, shows that
the powered Janus particles move superdiffusively. Previous experiments
and theoretical models~\cite{Zheng13,palacci_sedimentation_2010} have
shown that Janus particles move ballistically at short timescales (with
an exponent of 2) and diffusively at longer times scales (with an
exponent of 1). The non-Gaussianity we observe in the displacement
distributions is further evidence that the particles are in the
superdiffusive regime. Similar non-Gaussian behavior has been observed
and described theoretically by Zheng and coworkers~\cite{Zheng13} for
catalytic Janus particles in bulk. The origin of the non-Gaussian
behavior is the ballistic motion of the particles, which does not lead
to random-walk motion on these timescales. An exponent of 1.10 suggests
that our observations of the motion may be in a region of timescales
between the ballistic and diffusive regimes, but closer to the
diffusive.

At long timescales, we expect the diffusion to be enhanced. Because we
are close to this long-timescale diffusive regime, we fit the MSD to a
diffusion model, $\Delta r^2=4D\Delta t$, where $D$ is the apparent
diffusion coefficient. We find that for the unpowered Janus particles,
$D=0.3513 \pm 0.0002$ $\upmu$m$^2$/s. The value for the unpowered
particles is close to that measured for tracer particles in the absence
of fuel (see next subsection), as expected, though it is slightly lower
than the Stokes-Einstein estimate $D=0.48~\upmu$m$^2$/s for a
1-$\upmu$m-diameter particle in water at 25$^\circ$C, perhaps because
the actual diameter of the particles is larger than 1~$\upmu$m.

The diffusion coefficient for the powered Janus particles is
$D=0.5007\pm 0.0016$ $\upmu$m$^2$/s, more than 40\% higher than that for
the unpowered particles. This result suggests that, on the timescales
that we observe the motion, the particles are close to the enhanced
diffusion regime. We say ``close to'' because the exponent of 1.10 and
the displacement distributions in Figure~\ref{fig:superdiffusion}
indicate that there is still some ballistic motion on these timescales.

\subsection{Motion of Janus particles and tracers is weakly coupled}

\begin{figure}
\begin{center}
  \includegraphics{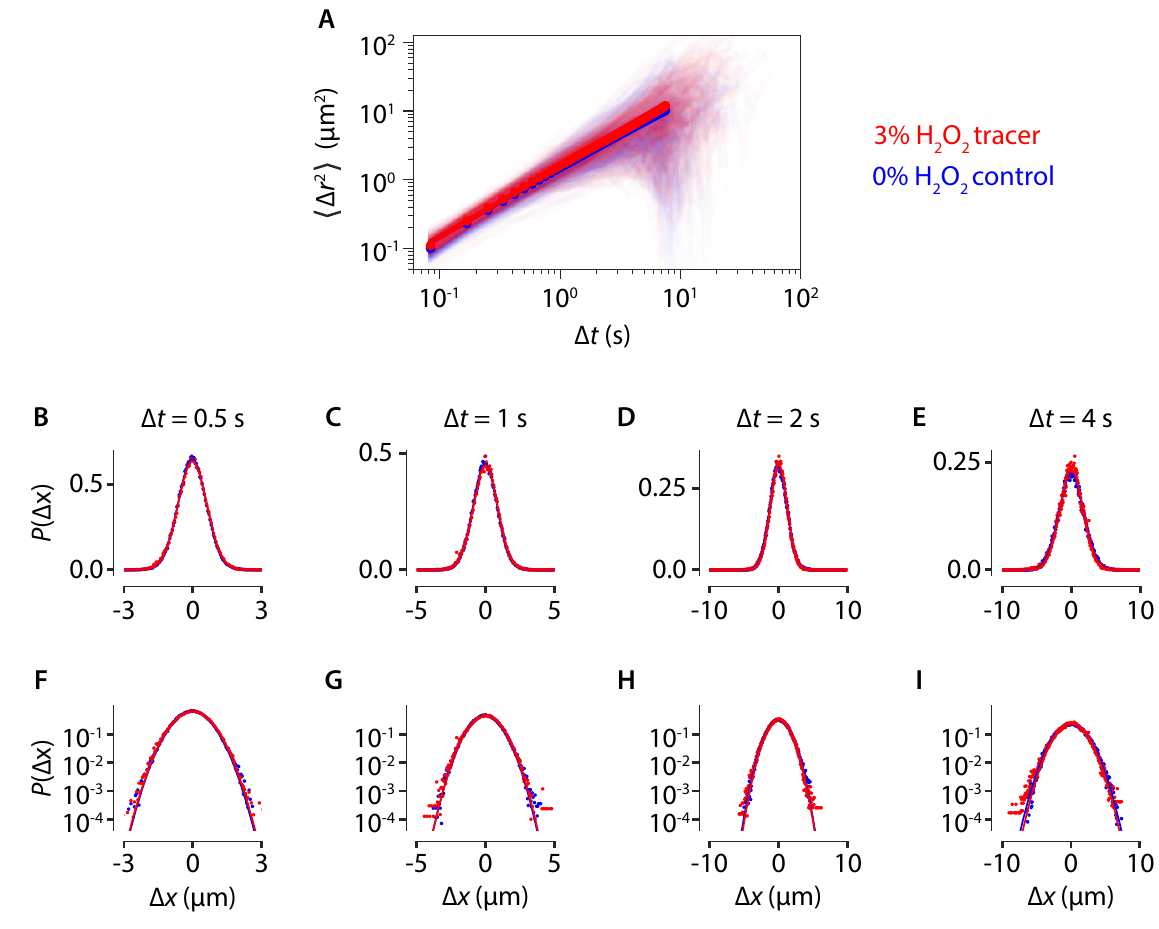}
  \caption{(Color online) The motion of tracer particles is weakly
    coupled to that of powered Janus particles. A: Ensemble mean square
    displacement (MSD) of the tracer particles as a function of lag
    time. Red data is for particles in the presence of Janus particles
    and hydrogen peroxide. Blue data is for particles in the presence of
    Janus particles but without hydrogen peroxide. The light traces show
    the MSD for individual particles, to indicate the spread of particle
    MSDs. B--E: Plots of the displacement distribution $P(\Delta x)$ of
    tracer particles in the presence of powered Janus particles (red)
    and in the presence of unpowered Janus particles (blue) Janus at lag
    times of 0.5~s (B), 1~s (C), 2~s (D), and 4~s (E). F--I: the same
    distributions shown on log-linear axes. The trajectories for these
    plots are drift-subtracted.}
  \label{fig:coupling}
\end{center}
\end{figure}

Having shown that the Janus particles are actively driven, even inside
the emulsion droplets, we now examine the coupling between the Janus
particles and the tracer particles. An analysis of the displacement
distributions (Figure~\ref{fig:coupling}) at different lag times shows
little difference between tracer particles in droplets with powered
Janus particles and tracer particles in droplets not containing fuel. In
both cases, the displacement distribution is Gaussian, and the widths of
the Gaussians with and without fuel are nearly identical. A fit to a
diffusion model, $\Delta r^2 =4D\Delta t$, where $D$ is the apparent
diffusion coefficient, yields $D=0.3880 \pm 0.0003$ $\upmu$m$^2$/s for
tracers in the presence of powered Janus particles, and
$D=0.3816\pm0.0002$ $\upmu$m$^2$/s for tracers in the presence of
unpowered Janus particles. Though there is a statistically significant
difference between the two diffusion coefficients, it is small (less
than 2\%). Thus, even though the Janus particles show enhanced
diffusion, the tracer particles show much smaller enhancement,
indicating that the motion of the two types of particles is weakly
coupled.

\section{Discussion}
\label{sec:discussion}

Our results show that Janus particles can exhibit enhanced diffusion
inside a confined system. The enhancement is evidenced by a 40\%
increase in the diffusion coefficient at lag times of approximately 1~s.
The exponent of the mean square displacement is approximately 1.10 at
these lag times, indicating that there is still some observable
ballistic motion at these timescales. In the established physical
picture of how active particles move~\cite{bechinger_active_2016}, the
enhanced diffusion regime arises from powered, ballistic motion at short
time scales combined with rotational diffusion at longer timescales,
which randomizes the direction of the ballistic motion.

Following Zheng and coworkers~\cite{Zheng13}, we calculate the timescale
at which we expect the transition from ballistic motion to enhanced
diffusion to begin. This timescale is $t=1/(2D_r)$, where $D_r$ is the
rotational diffusion coefficient, which is approximately 1~s$^{-1}$ for
our system. We find $t\approx0.5$ s, which is just below the range of
lag times (1--4~s) at which we observe the enhanced diffusion. Thus, the
observation of enhanced diffusion (and the slightly superdiffusive
exponent) make sense in the context of this theoretical framework: We
are observing the motion at timescales slightly larger than the
transition time, so that we see enhanced diffusion along with some
residual ballistic motion, which is apparent in the displacement
distributions.

However, the activity of the Janus particles does not seem to enhance
the diffusion of the tracers. There are several possible reasons for the
lack of coupling between the motion of the two types of particles.
First, the coupling between particles is hydrodynamic, and because the
Janus particles are active, they act as force
dipoles~\cite{bechinger_active_2016}. The velocity in the far field
therefore decays with distance $r$ as $1/r^2$, compared to $1/r$ for an
externally driven flow field~\cite{lauga_hydrodynamics_2009}. Therefore,
at the low volume fractions we use, the hydrodynamic coupling might be
weak. Furthermore, the magnitude of the force dipole might also be small
for our system, because we use low concentrations of hydrogen peroxide
in order to delay bubble formation.

Second, the efficiency of conversion of chemical energy to directed
motion is many orders of magnitude smaller for catalytic micromotors
such as our Janus particles than for molecular
motors~\cite{wang_understanding_2013}. A more efficient active particle
would likely show a larger force dipole and stronger coupling to the
tracer particles.

Third, in living cells the cytoplasm is crowded with proteins and
filaments, making it a weak elastic gel~\cite{Guo14}. In cells depleted
of ATP, passive tracer particles show very small mean square
displacements~\cite{Guo14}, indicating that they are trapped by the
elasticity of the medium. The enhanced diffusion of the tracer particles
in the presence of ATP-driven motor activity is therefore due to the
nonlinear elastic properties of the medium. By contrast, the fluid
inside the droplets in our system is purely viscous and linear.

These considerations suggest several future directions for making an
improved model system for mimicking transport inside the cytoplasm. For
example, one might increase the concentration of the motor particles.
However, a higher concentration of Janus particles would rapidly deplete
the hydrogen peroxide and increase the rate of bubble formation. An
alternative approach is to use biological motors instead of synthetic
ones, since their efficiency is orders of magnitude higher than that of
the Janus particles. Bacterial baths are a useful point of comparison:
many studies have shown that the diffusion of tracer particles is
significantly enhanced---by a factor of two or more---in
three-dimensional suspensions of motile \textit{E.\ coli} bacteria at
volume fractions of active particles comparable
to~\cite{kim_enhanced_2004,wilson_differential_2011} or even
smaller~\cite{chen_fluctuations_2007} than ours (1\%). The reason for
the stronger coupling may be that the efficiency, and hence the force
dipole, is much larger for a bacterium than for a catalytic Janus
particle.

If one wanted to use synthetic motors, light-driven
Janus particles~\cite{bechinger_active_2016} might be preferable, since
they circumvent the problem of bubble formation. However, the binary
liquid mixture needed for such particles might not be compatible with
the emulsion system we show here, and a different method of confinement
might be required.

With our current system of catalytic Janus particles, it would be
interesting to investigate the effects of crowding and elasticity of the
medium. Although the Janus particles have a weak effect on the motion of
tracer particles within a viscous medium, they might have a much larger
effect on particles in a weak elastic gel, because the thermally driven
diffusion of the tracer particles within the gel would be very small.

\section{Conclusion}
\label{sec:conclusion}

We have shown that active Janus particles, passive tracer particles, and
chemical fuel (hydrogen peroxide) can be confined within emulsion
droplets, such that the Janus particles show enhanced diffusion and do
not stick to the droplet interfaces. In these experiments we are careful
to correct for bulk rotational and translational motion induced by
bubble formation. After this correction we find that the motion of the
passive tracer particles is weakly coupled to that of the active Janus
particles, likely because the efficiency and concentration of the Janus
particles is too low. Nonetheless, the results point the way forward
toward a model system that can be used to understand active transport in
living cells, where the motion of motors can significantly enhance the
diffusion of unattached particles within the cytoplasm. Future versions
of such a system might benefit from introducing crowding agents or more
efficient active particles. If diffusion-limited chemical reactions can
be sped up by the particles, it might enable the use of artificial cells
as biochemical factories.

\pdfbookmark{Acknowledgments}{Acknowledgments}
\section*{Acknowledgments}

We acknowledge support from the Army Research Office through the MURI
program under award no.\ W911NF-13-1-0383, from the Harvard Materials
Research Science and Engineering Center through NSF grant no.\
DMR-1420570, and from the Swedish Research Council (Vetenskapsr{\aa}det)
637-2013-414. We thank Ilona Kretzschmar, Sepideh Razavi, and Bin Ren
for guidance in making Janus particles and W.\ Benjamin Rogers and
Kinneret Keren for helpful discussions. This work was performed in part
at the Harvard Center for Nanoscale Systems (CNS), a member of the
National Nanotechnology Infrastructure Network (NNIN), which is
supported by the NSF (ECS-1541959).

\pdfbookmark{Author contributions}{Author contributions}
\section*{Author contributions}

V.R.H., T.G.D., and V.N.M.\ conceived of the presented idea. Z.C.C.\
prepared samples and collected data. V.R.H., Z.C.C., and V.N.M.\
performed analysis. \.{I}.G.\ developed the method for making Janus
particles and contributed diagrams. V.R.H., T.G.D., \.{I}.G., and
V.N.M.\ wrote the paper. All authors gave final approval for the paper
to be published.

\pdfbookmark{Data availability}{Data availability}
\section*{Data availability}

The datasets generated during the current study are available from the
corresponding author on reasonable request.

\pdfbookmark{References}{References}

\end{document}